\documentclass{elsart}
\usepackage{graphicx}
\usepackage{epsfig}
\usepackage{amssymb}

\begin{document}

\begin{frontmatter}

\title{Reaction-noise induced homochirality}

\author[CAB]{David Hochberg\corauthref{cor}},
\corauth[cor]{Corresponding author.} \ead{hochberg@laeff.esa.es}
\author[CAB]{Mar\'{i}a-Paz Zorzano},
\ead{zorzanomm@inta.es}

\address[CAB]{Centro de Astrobiolog\'{\i}a (CSIC-INTA), Ctra. Ajalvir
Km. 4, 28850 Torrej\'{o}n de Ardoz, Madrid, Spain}
\begin{abstract}
Starting from the chemical master equation, we employ field
theoretic techniques to derive Langevin-type equations that exactly
describe the stochastic dynamics of the Frank chiral amplification
model with spatial diffusion. The intrinsic multiplicative noise
properties are completely and rigorously derived by this procedure.
We carry out numerical simulations in two spatial dimensions. When
the inherent spatio-temporal fluctuations are properly included,
then complete chiral amplification results from a purely racemic
initial configuration. Phase separation can also arise in which the
enantiomers coexist in spatially segregated domains separated by a
sharp racemic interface or boundary.
\end{abstract}

\end{frontmatter}

\section{\label{sec:intro} Introduction}

Questions about the origin of life have always held great
fascination. One especially important problem is to explain chiral
symmetry breaking in nature, why for example, it came to be that the
nucleotide links of RNA and DNA incorporate exclusively
dextro-rotary (D) ribose and D-deoxyribose while the enzymes involve
only laevo-rotary (L) enantiomers of amino acids. Mirror symmetry is
broken in the bioorganic world and life as we know it is invariably
linked with homochirality. For recent reviews that survey existing
hypotheses concerning this phenomenon, specific experimental
realizations and additional references, see
\cite{Avet,Avalos2,Kondepudi}.

A simple chemical scheme to explain spontaneous generation of chiral
asymmetry was proposed by Frank in 1953 \cite{Frank}. In it original
version, the Frank model leads to an amplified production of a
chiral enantiomer from a racemic mixture seeded with an initial
bias. If however the initial state of the system is strictly
racemic, it remains racemic for all time. The problem is to explain
the origin of the initial chiral seed. The purpose of this Letter is
to demonstrate that when the inherent spatio-temporal fluctuations
are properly included, then complete chiral amplification results
from a purely racemic initial configuration. Phase separation can
also arise in which the enantiomers coexist in spatially segregated
domains separated by a racemic interface or boundary.  The internal
fluctuations responsible for these two a-priori unpredictable
outcomes are a natural consequence of the fact that chemical
reactions are generally diffusion-limited, and take place in
imperfectly mixed systems \cite{Epstein}.

The general method we employ in this Letter is best appreciated by
thoroughly working through a simple example based on an extension of
the Frank model, which we now introduce. Let L and D denote a pair
of enantiomers. Then the modified Frank model we will study is
described by the following reaction scheme \cite{GTV}:

Autocatalytic production:
\begin{equation}\label{autoLD}
\textrm{L} + \textrm{A} \stackrel{k_1}{\rightleftharpoons
\atop{\small k_3}}\textrm{L} + \textrm{L}, \qquad \textrm{D} +
\textrm{A} \stackrel{k_1} {\rightleftharpoons \atop{\small k_3}}
\textrm{D} + \textrm{D}
\end{equation}
Mutual destruction, or dimerization, in a second order reaction:
\begin{equation}\label{mutual}
\textrm{L} + \textrm{D} \stackrel{k_2}{\longrightarrow} \textrm{P}.
\end{equation}
The $k_i$ denote the rate constants and we take the achiral
substance A as a uniform constant background. The difference between
this and the original Frank model\footnote{For the direct formation
of chiral matter from achiral substrate, we would have had to
include the steps $\textrm{A} \rightarrow \textrm{L}(k_0)$ and
$\textrm{A} \rightarrow \textrm{D}(k_0)$. According to \cite{Buhse},
$k_0$ must be kept sufficiently small. Otherwise, these chirally
unspecific reactions would proceed too rapidly thus generating large
amounts of racemic matter that swamps the amplification process
driven by the autocatalytic (\ref{autoLD}) and mutual inhibition
steps (\ref{mutual}). To take this important observation into
account, we simply set $k_0 = 0$ from the outset.} lies in the
open-flow reactor nature of the process and the fact that the
reaction (1) is allowed to be reversible ($k_3 \geq 0$)
\cite{Mason}. The system is fed by an input of the achiral substrate
A, whereas the output consists of the inactive product P, and the
excess of the enantiomers. We further assume that each enantiomer
diffuses with the same diffusion constant $D$ and include this
feature in the master equation description of this process.

\section{\label{sec:field} Effective Bosonic Field Theory}

Our starting point for a systematic treatment of the above reaction
scheme is an appropriate master equation. On a microscopic level,
this comprises an exact description of the dynamics. From this
equation, it is then straightforward to derive an effective
stochastic field theory. The salient steps are given in detail
below.

The chemical master equation for the kinetic scheme in
Eqs.(\ref{autoLD},\ref{mutual}) is mapped to a ``second-quantized"
description following Doi \cite{Doi}. Briefly, we introduce
annihilation and creation operators $a_i$ and $a^\dag_i$ for L,
$b_i$ and $b^\dag_i$ for D at each lattice site $i$, obeying the
commutation relations $[a_i,a_j^{\dag}] = \delta_{ij}$,
$[b_i,b_j^{\dag}] = \delta_{ij}$. The vacuum state (corresponding to
the configuration containing zero particles) satisfies $a_i|0\rangle
= b_i|0\rangle = 0$. Furthermore, $a_i|m_i\rangle =
m_i|m_i-1\rangle, a_i^{\dag}|m_i\rangle = |m_i+1\rangle$ and
$a^{\dag}_ia_i|m_i\rangle = m_i|m_i\rangle$ (and similarly for the
$b$-sector). We then define the time-dependent state vector
\begin{equation}\label{wavefunction}
|\Psi(t)\rangle = \sum_{\{m\},\{n\}}P(\{m\},\{n\},t) \prod_i(
{a}_i^\dag)^{m_i}( {b}_i^\dag)^{n_i}|0\rangle,
\end{equation}
where $P(\{m\},\{n\},t)$ is the probability distribution to find
$m_i,n_i$ particles of type L,D, respectively, at each site $i$, and
satisfies the master equation.
Then by means of Eq.(\ref{wavefunction}), the master equation can be
written as a ``Schr\"{o}dinger" equation
\begin{equation}
\label{schrodinger} -\frac{\partial |\Psi(t)\rangle}{\partial t} =
 {H}|\Psi(t)\rangle, \qquad \Rightarrow \qquad |\Psi(t)\rangle =
 \exp\big(-H t\big)|\Psi(0)\rangle,
\end{equation}
where the lattice hamiltonian $H$ is calculated to be
\begin{eqnarray}\label{hamiltonian}
H &=& \frac{D}{l^2}\sum_{(i,j)} \big(a_i^{\dag} - a_j^{\dag}\big)
\big(a_i - a_j\big) + \frac{D}{l^2}\sum_{(i,j)} \big(b_i^{\dag} -
b_j^{\dag}\big) \big(b_i - b_j\big) \nonumber \\
&-& \sum_i \left\{ k_1\big((a^{\dag}_i)^2a_i - a^{\dag}_ia_i \big) +
k_1\big((b^{\dag}_i)^2b_i - b^{\dag}_ib_i \big) \right. \nonumber \\
&+& k_3\big(a^{\dag}_ia_i^2 - (a^{\dag}_i)^2a_i^2 \big) +
k_3\big(b^{\dag}_ib_i^2
- (b^{\dag}_i)^2b_i^2 \big) \nonumber \\
&+& \left. k_2\big(a_ib_i - a^{\dag}_ia_ib^{\dag}_ib_i \big)
\right\},
\end{eqnarray}
and $(i,j)$ denotes the sum over all lattice sites $i$ (with lattice
spacing $l$) and their nearest neighbors $j$ in a $d$-dimensional
space.

Now take the continuum limit ($l \rightarrow 0$) \cite{Peliti} in
order to obtain the path integral representation of the underlying
stochastic dynamics. Thus we can write the time-evolution operator
\begin{equation}
\exp\big(-H t\big) = \int \mathcal{D}a\mathcal{D}\bar a\mathcal{D}b
\mathcal{D}\bar b \exp(-S[a,\bar a,b,\bar b])
\end{equation}
in terms of continuous fields $a(\mathbf{x},t),\bar
a(\mathbf{x},t),b(\mathbf{x},t),\bar b(\mathbf{x},t)$ with weight
$\exp(-S[a,\bar a,b,\bar b])$. Next, perform the shift $\bar a = 1 +
a^*$, $\bar b = 1 + b^*$ on $S$. This yields the shifted action for
any space dimension $d$, which is found to be:
\begin{eqnarray}\label{actionahift2}
S[a,a^*,b,b^*] &=& \int d^d\mathbf{x} \int dt \, \left\{ a^*\Big(
\partial_ta -D \nabla^2 a+ k_2 ab -k_1 a + k_3a^2 \Big) \right.
\nonumber \\
&+&  b^*\Big(
\partial_tb -D \nabla^2 b+ k_2 ab -k_1 b + k_3b^2 \Big) +
{a^*}^2\big(k_3a^2 - k_1a \big) \nonumber \\
&+& \left. {b^*}^2\big(k_3b^2 - k_1b \big) + k_2a^*b^*a b \right\}.
\end{eqnarray}
The action $S$ contains \textit{all the dynamics} of the reaction
scheme (\ref{autoLD},\ref{mutual}) including diffusion,  and apart
from taking the continuum limit, is exact. In particular, \textit{no
assumptions} regarding the form of the noise are required. At this
stage, there are at least two independent ways to extract detailed
information from $S$: by (i) employing field theoretic
renormalization group techniques \cite{THVL} or (ii) by exploiting
the exact equivalence of this action to coupled Langevin equations.
The former is suited for understanding chiral symmetry breaking as a
problem in dynamic critical phenomena, the latter is ideal for
understanding chiral amplification, evolution of enantiomeric
excesses and net reactor yield as a problem in spatial pattern
formation.

\section{\label{sec:langevin} The Exact Langevin Equations}

For the final step we use the gaussian transformation \cite{Amit}
which allows us to integrate exactly over the conjugate fields
$a^*,b^*$, appearing in the path integral $\int \mathcal{D}a
\mathcal{D}{a^*}\mathcal{D}b \mathcal{D}{b^*} \,
e^{-S[a,a^*,b,b^*]}$. This final step yields a product of
delta-functional constraints which in turn, immediately imply the
following set of \textit{exact} coupled stochastic partial
differential equations:
\begin{eqnarray}\label{a}
\partial_t a &=& D\nabla^2 a + k_1a - k_2 ab - k_3a^2 +
\eta_a(\mathbf{x},t), \\ \label{b}
\partial_t b &=& D\nabla^2 b + k_1b - k_2 ab - k_3b^2 +
\eta_b(\mathbf{x},t),
\end{eqnarray}
where the intrinsic multiplicative reaction noise is
\textit{completely and exactly} specified as follows, namely
$\langle \eta_a \rangle = \langle \eta_b \rangle = 0$ and
\begin{eqnarray}\label{noise}
\langle \eta_a (\mathbf{x},t)  \eta_a (\mathbf{x'},t') \rangle &=&
(k_1 - k_3 a(\mathbf{x},t))a(\mathbf{x},t) \,\delta^d(\mathbf{x}-\mathbf{x'})\delta(t - t'),\\
\langle \eta_b (\mathbf{x},t)  \eta_b (\mathbf{x'},t') \rangle &=&
(k_1 - k_3 b(\mathbf{x},t))b(\mathbf{x},t) \,\delta^d(\mathbf{x}-\mathbf{x'})\delta(t - t'),\\
\langle \eta_a (\mathbf{x},t)  \eta_b (\mathbf{x'},t') \rangle &=&
-\frac{1}{2}k_2 a(\mathbf{x},t)b(\mathbf{x},t) \,
\delta^d(\mathbf{x}-\mathbf{x'})\delta(t - t').
\end{eqnarray}
In the absence of noise (mean field limit) the fields
$a(\mathbf{x},t),b(\mathbf{x},t)$ correspond to the coarse-grained
local densities of the L,D particles. With noise, these fields are
generally complex, and do not represent the physical densities.
However, the averages $\langle a(\mathbf{x},t)\rangle,\langle
b(\mathbf{x},t)\rangle$ are real and do correspond to particle
densities \cite{HT}.

For convenience, we cast this system in terms of nondimensional
couplings and fields. Introducing $L$ and $T$ for spatial and
temporal dimensions, dimensional analysis implies that $[a] = [b] =
1/L^d, \,[D] = L^2/T, \, [k_1] = T^{-1}, \, [k_2] =[k_3] = L^d/T, \,
[\eta] = 1/{TL^d}$. So, define dimensionless fields $\tilde a =
(k_2/k_1) a$, $\tilde b = (k_2/k_1) b$, time $\tau = k_1 t$,
coordinates $\hat x_j = \sqrt{k_1/D}\, x_j$ and dimensionless noises
$\tilde \eta_a = (k_2/k_1^2) \eta_a$, $\tilde \eta_b = (k_2/k_1^2)
\eta_b$. Then the equations Eqs.(\ref{a},\ref{b}) can be written as
\begin{eqnarray}\label{aa}
\partial_{\tau} \tilde a &=& {\hat \nabla^2} {\tilde a} + \tilde a - {\tilde a}{\tilde b} -
g {\tilde a}^2 + \tilde \eta_a, \\
\label{bb}
\partial_{\tau} \tilde b &=& {\hat \nabla^2} {\tilde b} + \tilde b - {\tilde a}{\tilde b} -
g {\tilde b}^2 + \tilde \eta_b,
\end{eqnarray}
and the correlation matrix of the dimensionless noises is given by
\begin{equation}\label{noisecorrel}
B = \epsilon \, \left(
  \begin{array}{cc}
    \tilde a(1 - g \tilde a) & -\frac{1}{2}\tilde a \tilde b \\
                                          & \\
    -\frac{1}{2}{\tilde a}{\tilde b} & \tilde b(1 - g \tilde b) \\
  \end{array}
\right),
\end{equation}
where $g=\frac{k_3}{k_2}$ and $\epsilon =
\frac{k_2}{k_1}\Big(\frac{k_1}{D}\Big)^{d/2}$ is the spatial
dimension-dependent noise amplitude. Note that in $d=2$ this reduces
to $\epsilon = \frac{k_2}{D}$, indicating that the noise is
controlled via a competition between the dimerization reaction and
spatial diffusion.

We apply the Cholesky decomposition $B=M M^T$ to extract the square
root of the matrix in the form of a lower triangular matrix $M$
\cite{HZM}. We will use this decomposition to relate the noise to a
new real white Gaussian noise $\vec{\xi}= (\xi_1,\xi_2)$, $\langle
\xi_i(\hat \mathbf{x},\tau)\xi_j(\hat \mathbf{x}',\tau')\rangle =
 \delta_{ij}\delta(\hat \mathbf{x}-\hat \mathbf{x}')\delta(\tau-\tau')$,
such that $(\tilde \eta_a,\tilde\eta_b) =
\vec{\tilde\eta}=M\vec{\xi}$ (and thus $\vec{\tilde
\eta}^T=\vec{\xi}^TM^T$). Notice that by doing so the condition
$<\vec{\tilde \eta}\vec{\tilde \eta}'^T>=<M\vec{\xi}\vec{\xi}'^TM^T
>=M<\vec{\xi}\vec{\xi}'^T>M^T=M M^T=B$ is satisfied. This
decomposition will allow us to separate the real and imaginary parts
of the noise, a useful feature to have for setting up a numerical
integration of the (complex-valued) stochastic reaction-diffusion
system \cite{HZM}. We verify that
\begin{equation}
M = \, \left(
  \begin{array}{cc}
    \sqrt{B_{11}} &  0 \\
                  & \\
    \frac{B_{21}}{\sqrt{B_{11}}} & \sqrt{ B_{22}-\frac{B_{21}^2}{B_{11}}} \\
  \end{array}
\right)
\end{equation}
\begin{equation}
 = \sqrt{\epsilon}\,\left(
      \begin{array}{cc}
        \sqrt{\tilde a}\big(1 - g\tilde a \big)^{1/2} &  0 \\
          & \\
        -\frac{1}{2}\tilde b \sqrt{\tilde a} \big(1 - g\tilde a \big)^{-1/2}
        & \,\,
        \big(1 - g\tilde a \big)^{-1/2}\sqrt{\tilde b}
        \sqrt{(1- g\tilde b)(1- g \tilde a) - \frac{1}{4}\tilde a \tilde b}
      \end{array}
    \right)
\end{equation}
yields the desired decomposition of $B$.

\section{\label{sec:num} Numerical results}

The non-trivial mean-field solutions of Eqs. (\ref{aa},\ref{bb}), in
the absence of fluctuations (deterministic case, $\epsilon=0$), have
been worked out in \cite{GTV}:
\begin{itemize}
\item{(I) if $g<1$: $\tilde{b}=0,\ \tilde{a}=\frac{1}{g}$, homochiral
absorbing state solution},
\item{(II) if $g<1$: $\tilde{a}=0,\ \tilde{b}=\frac{1}{g}$, homochiral
absorbing state solution},
\item{(III) if $g>1$: $\tilde{a}=\tilde{b}=\frac{1}{1+g}$ racemic
active state solution},
\end{itemize}
where $g=\frac{k_3}{k_2}$. In solutions (I) and (II), one of the two
enantiomers is amplified to its maximal value and the other one
disappears, this corresponds to pure chiral amplification. On the
contrary, in solution (III) the two enantiomers coexist with the
same concentration, this corresponds to the racemic solution. This
scenario is only valid in the mean-field limit. If the full problem
is treated, then the noise due to spatial fluctuations and the
diffusion of reactants must properly be taken into account.

We will solve the full stochastic two-dimensional version of Eqs.
(\ref{aa},\ref{bb}) numerically, using reflecting boundary
conditions and a finite difference scheme with $\Delta \tau=0.05$,
$\Delta \hat{x}=\Delta \hat{y}=0.51$, and a grid of size $L \times L
=154\times154$ \footnote{In our numerical studies, the system size
did not have any effect on the stationary solutions, only on the
time required to reach equilibrium.}.

\begin{figure}[h]
\begin{center}
\begin{tabular}{ccccccc}
$\tilde{a}$ & \includegraphics[width=0.15
\textwidth]{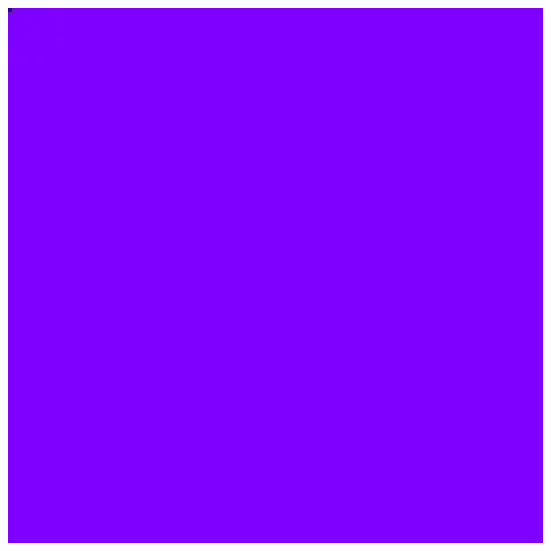}&
\includegraphics[width=0.15 \textwidth]{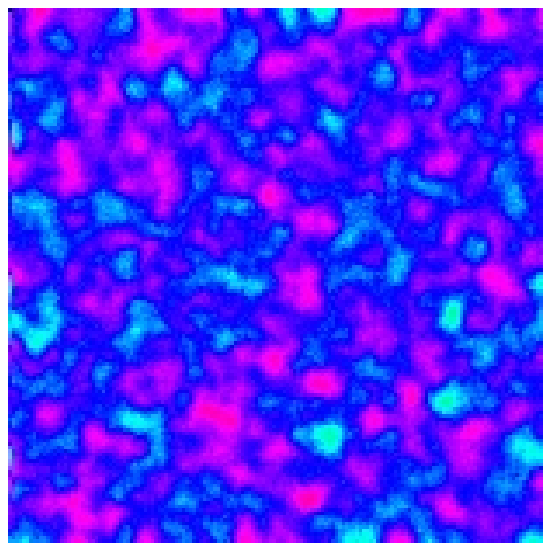}&
\includegraphics[width=0.15\textwidth]{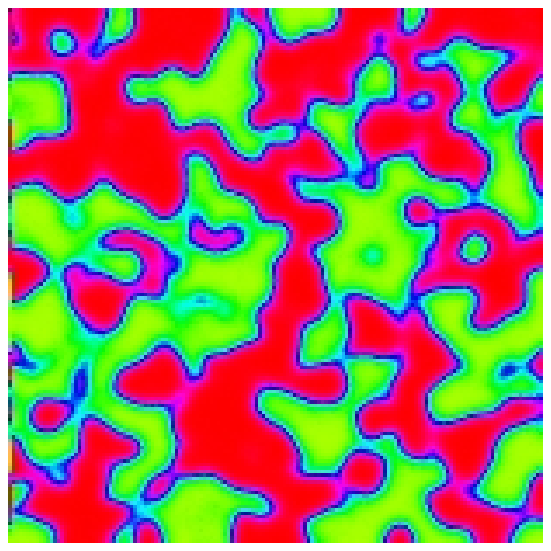}&
\includegraphics[width=0.15\textwidth]{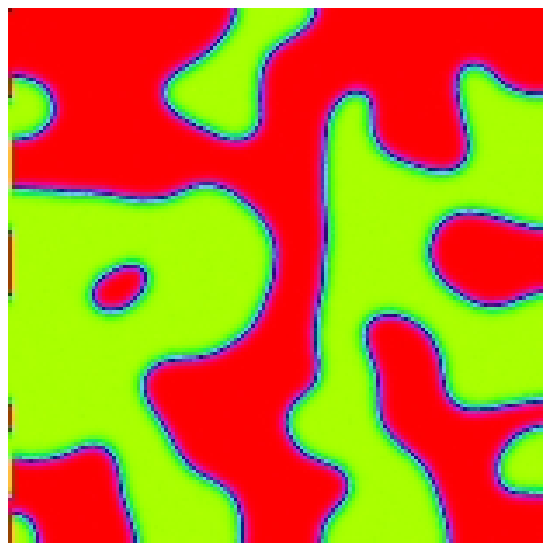}&
\includegraphics[width=0.15\textwidth]{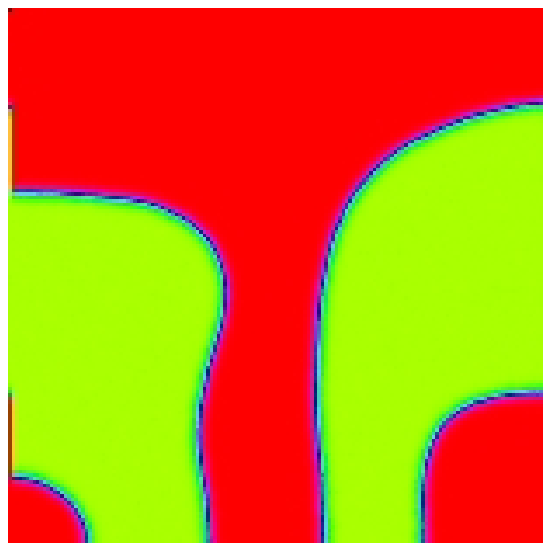}&
\includegraphics[width=0.15\textwidth]{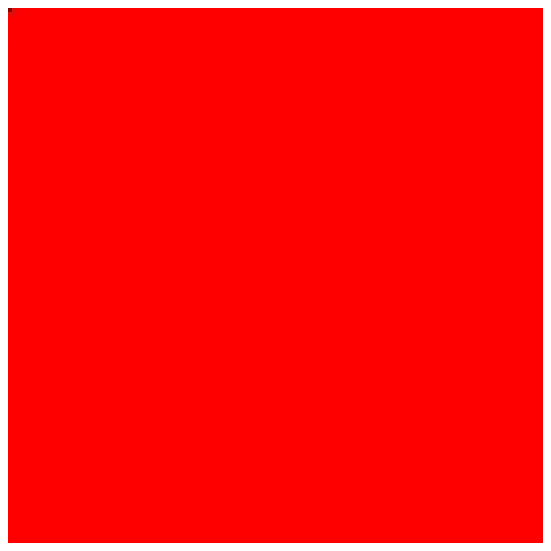}\\
$\tilde {b}$ &
\includegraphics[width=0.15 \textwidth]{InitialCondition.eps}&
\includegraphics[width=0.15 \textwidth]{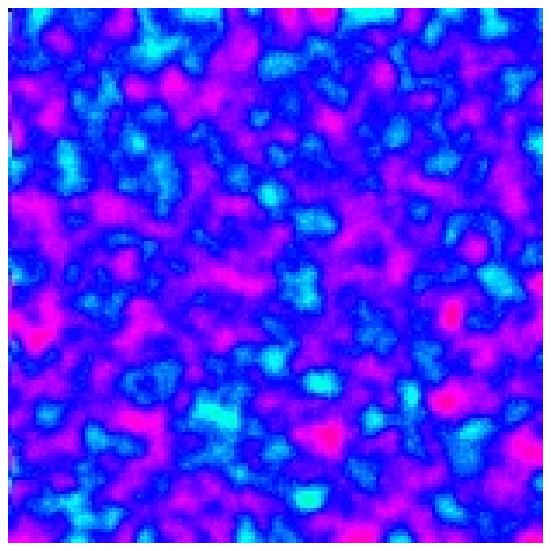}&
\includegraphics[width=0.15\textwidth]{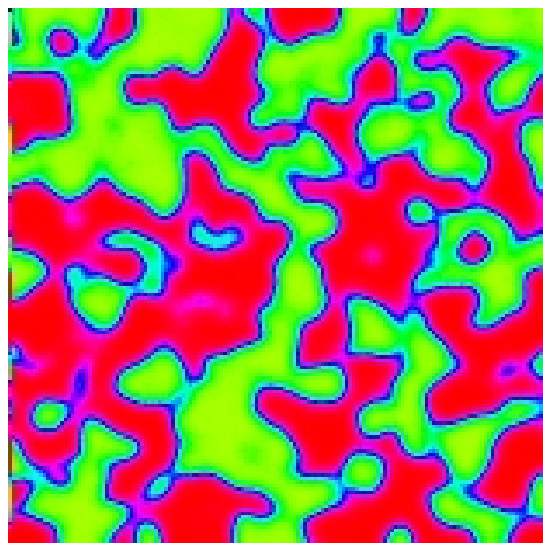}&
\includegraphics[width=0.15\textwidth]{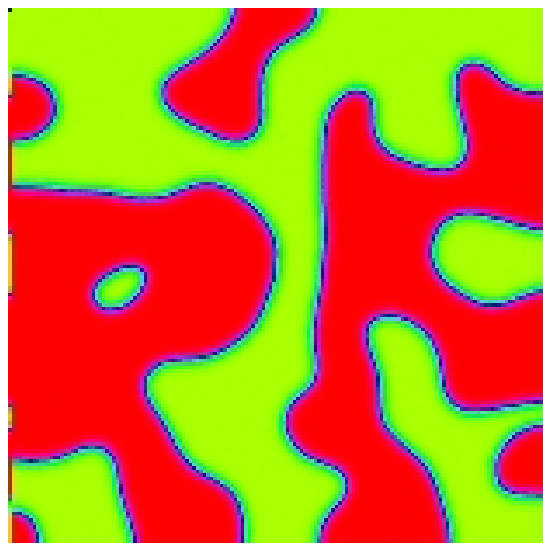}&
\includegraphics[width=0.15\textwidth]{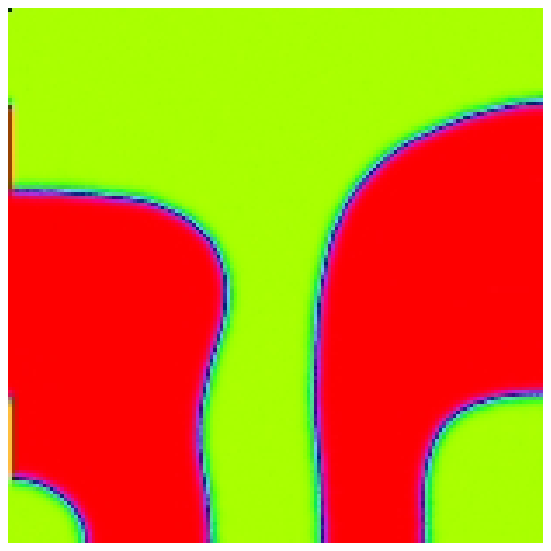}&
\includegraphics[width=0.15\textwidth]{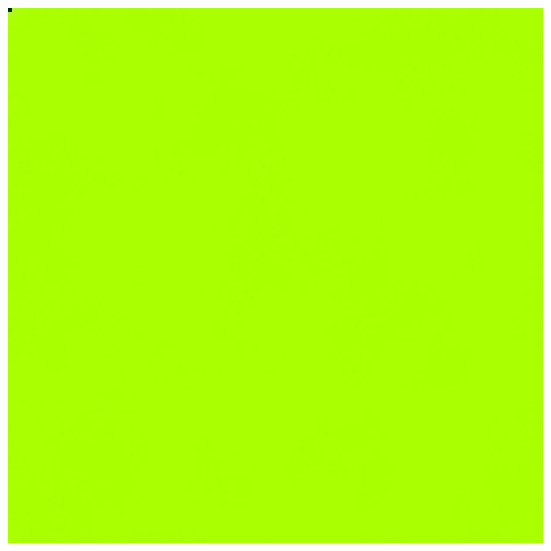}\\
\end{tabular}
\caption{\label{Exp1}Simulation 1. Starting from a homogeneous
racemic condition, the local random fluctuations induced by
imperfect mixing drive the system to a homochiral solution. Time
runs from left to right. See the text for details.}
\end{center}
\begin{center}
\begin{tabular}{ccccccc}
$\tilde{a}$ & \includegraphics[width=0.15
\textwidth]{InitialCondition.eps}&
\includegraphics[width=0.15 \textwidth]{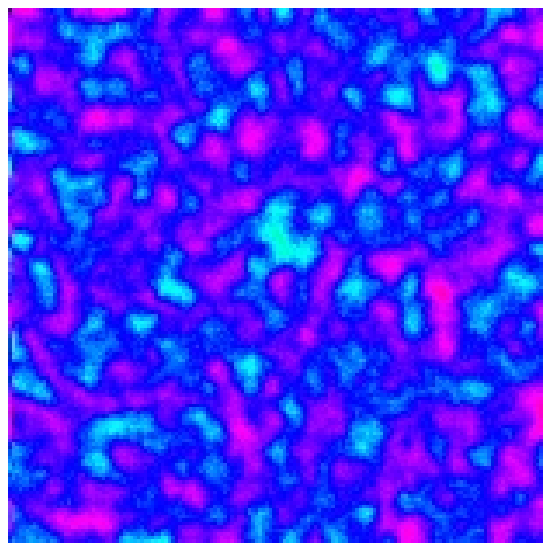}&
\includegraphics[width=0.15\textwidth]{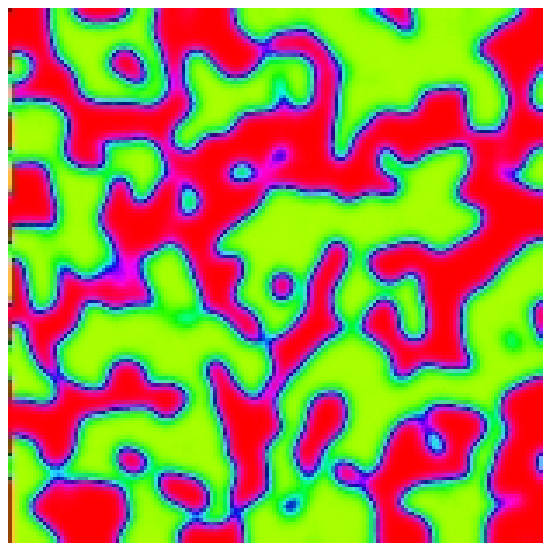}&
\includegraphics[width=0.15\textwidth]{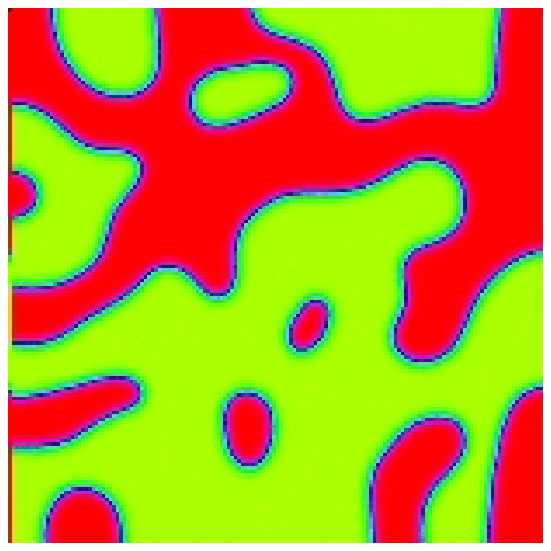}&
\includegraphics[width=0.15\textwidth]{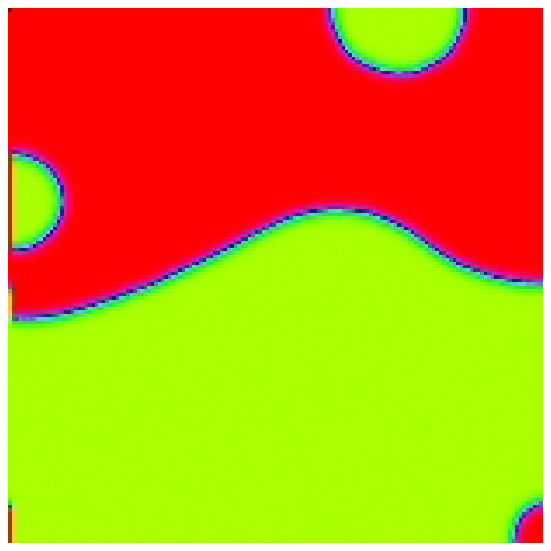}&
\includegraphics[width=0.15\textwidth]{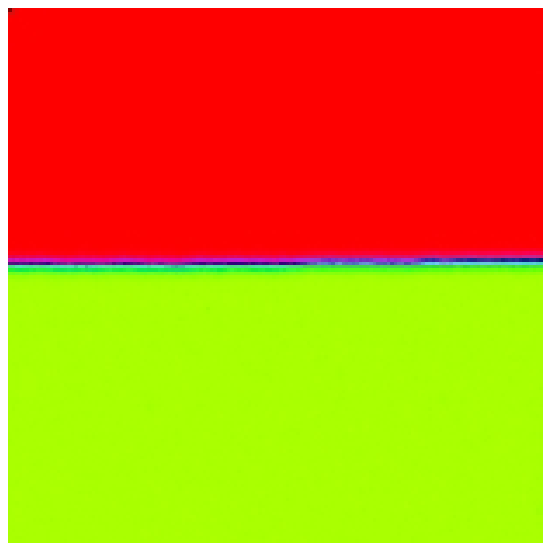}\\
$\tilde{b}$ & \includegraphics[width=0.15
\textwidth]{InitialCondition.eps}&
\includegraphics[width=0.15 \textwidth]{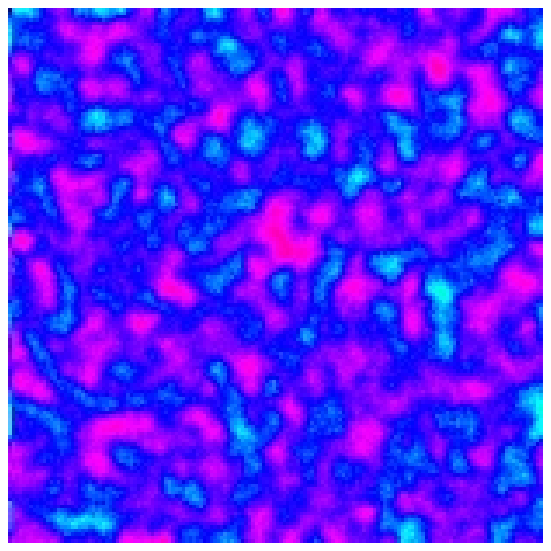}&
\includegraphics[width=0.15\textwidth]{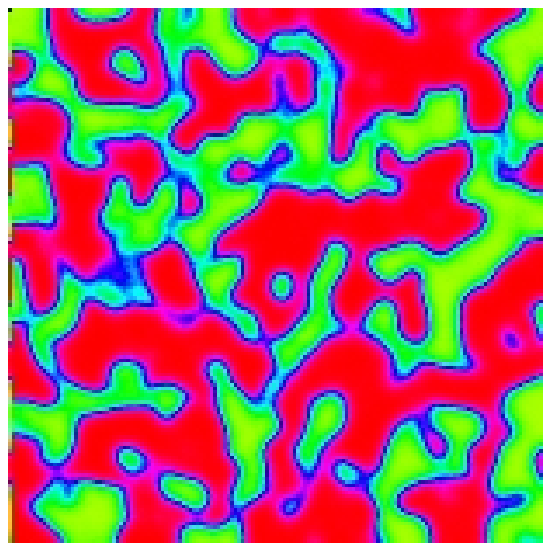}&
\includegraphics[width=0.15\textwidth]{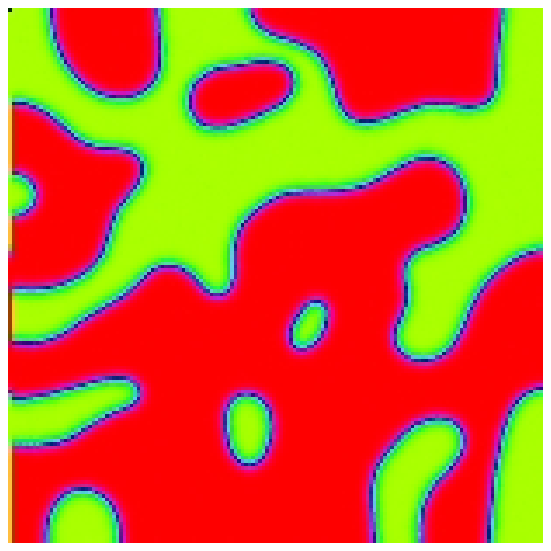}&
\includegraphics[width=0.15\textwidth]{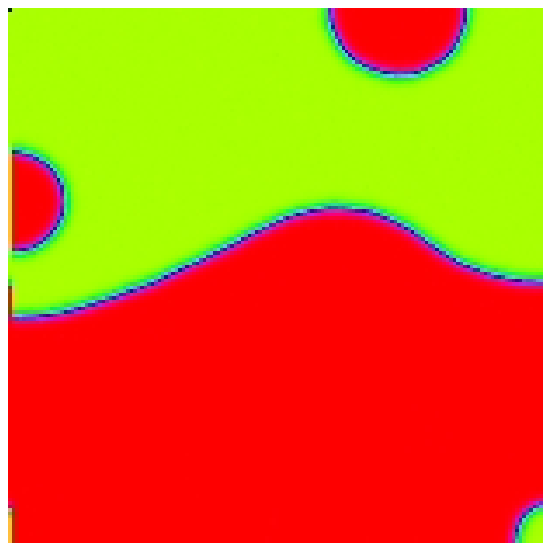}&
\includegraphics[width=0.15\textwidth]{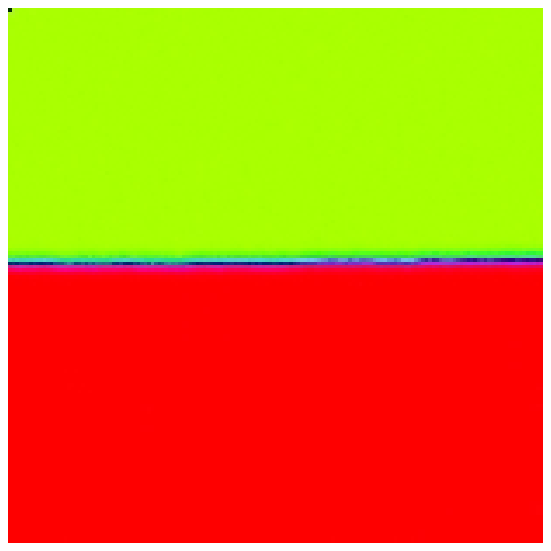}\\
\end{tabular}
\caption{\label{Exp2}Simulation 2. The system is solved again with
the same conditions as in Fig. \ref{Exp1}. The {\em new} random
fluctuations induced by imperfect mixing lead the system to a
different final state with spatial segregation. Time runs from left
to right.}
\end{center}
\end{figure}

We first considered the evolution of a system with
$g=\frac{0.09}{0.14}=0.643<1$. Two representative results are shown
in Figs. \ref{Exp1} and \ref{Exp2}. The initial condition was set to
a homogeneous, racemic situation
$(\tilde{a}_0=0.5,\tilde{b}_0=0.5)$, and the noise strength
$\epsilon=\frac{k_2}{D}=10^{-3}$ was fixed. A sequence of
simulations were independently executed: every time the same
mathematical problem was solved, the only difference being the
random sequence of applied noises, and the resulting evolution of
each realization. In the two sets of rows, we plot the time
evolution of the spatial distribution of the real part of
$\tilde{a}(\hat{x},\hat{y},\tau)$ (upper row) and
$\tilde{b}(\hat{x},\hat{y},\tau)$ (lower row). From left to right,
the spatial distributions are shown at $\tau=0,83,166,1111,8333$ and
$140000$, respectively. When displayed in color, red represents a
concentration of 0, purple represents a concentration of roughly
0.5, dark blue roughly 0.6 and green is the maximal concentration,
1.555 (=$\frac{1}{g}$).  In both simulations, after the transient
time ($\tau > 100$, i.e. after the first two frames of the sequences
in  Figs. \ref{Exp1} and \ref{Exp2}), we found that even if the
initial state was homogeneous and racemic (see purple squares in
Figures 1 and 2), the spatial fluctuations rapidly drive the system
to a pattern with marked spatial compartamentalization, where,
locally, the system is pure in one of the enantiomers \footnote{Thus
from the third square onward from left to right in both Figures 1
and 2, we see the formation of red and green zones. Top row Figure
1, red and green signal the absence of and maximal concentration of
$\tilde a$, respectively. Bottom row Figure 1: red and green signal
absence of and maximal concentration of $\tilde b$, respectively.
Final state in this sequence is zero concentration $\tilde a$ (red
square top row) and maximal concentration for $\tilde b$ (green
square bottom row). In Figure 2, the final state (right-hand most
squares in top and bottom rows) represents half the domain filled by
maximal $\tilde a$ and the complementary half square filled by
maximal $\tilde b$ concentrations, respectively, each half domain
separated by a racemic boundary line.}. This means that we can find
coexistence of local environments with the homochiral solutions (I)
or (II). In the boundaries between regions of different chiralities
there are racemic fronts of type (III), which in the mean-field
limit would have only been expected for $g>1$. Note the fully
complementary nature between the spatial patterns formed of the
$\tilde a$ and $\tilde b$-sequences during their evolution. Once the
system has reached this state, the different regions compete and
merge and the system evolves to one of the following final states:
\begin{itemize}
\item{ A) A homochiral, homogenous solution  where one of the enantiomers
has entirely vanished, see Fig. \ref{Exp1}.}
\item{B) A mixed state with strong spatial segregation of two immiscible
phases, where locally in each phase, one of the homochiral solutions
is realized, separated by planar racemic interface, see Fig. \ref{Exp2}.}
\end{itemize}
Notice that when the solution is homochiral (solutions (I) and (II))
the noise vanishes since the correlation matrix, Eq.
\ref{noisecorrel}, vanishes as well. Thus this final state is
absorbing. The racemic fronts are fluctuating and active.

\begin{figure}[h]
\begin{center}
\includegraphics[width=0.7 \textwidth]{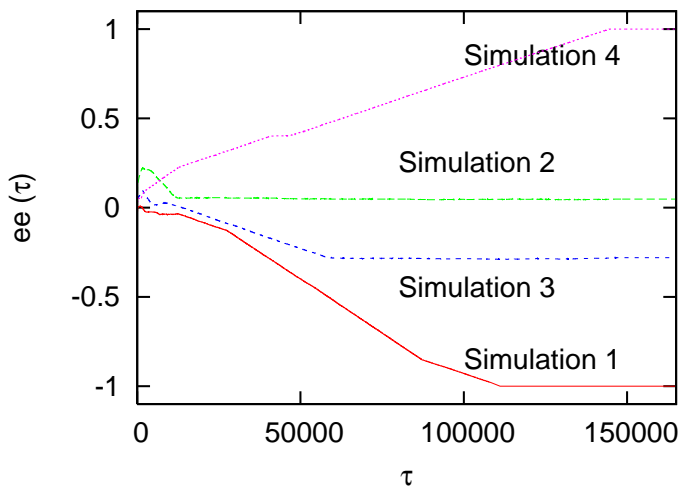}
\caption{\label{fig2} Time evolution of the spatially averaged
enantiomeric excess $ee (\tau)=\frac{<\tilde a> -<\tilde b>}{<\tilde
a>+<\tilde b>}$ of four representative simulations of the same study
case, starting from racemic conditions (not resolvable in the
figure). Due to the inherent symmetry of the reaction, either
enantiomer $\tilde a$ or $\tilde b$ can dominate. The final result
varies from one simulation to another. One can have total,
homogeneous, chiral amplification (Simulations 1 and 4) or local
chiral amplification with a racemic front (Simulations 2 and 3).}
\end{center}
\end{figure}

We verified that the complex solutions of Eqs.(\ref{aa},\ref{bb}),
once averaged over the fluctuations, do indeed yield real results,
in accord with the theoretical expectations \cite{HT}. Next, we
evaluate the time evolution of the spatially averaged (over
$\hat{x},\hat{y}$) densities for the two cases shown above and for
two other simulations representative of the inverse situation (i.e.,
where the system is dominated by the other enantiomer). We represent
in Fig. \ref{fig2} the time evolution of the averaged enantiomeric
excess $ee (\tau)=\frac{<\tilde a> - <\tilde b>}{<\tilde a> +
<\tilde b>}$. When the solution is spatially homogeneous with
$\tilde a$ at its maximal value, then $ee=+1$, for the inverse
situation where $\tilde b$ is amplified maximally, then $ee=-1$,
whereas for the intermediate situation with spatial
compartmentalization $0<ee < 1$ if the region with $\tilde a$ is
greater than the one with $\tilde b$, and  $-1<ee < 0$ for the
opposite situation. All these outcomes were found with roughly equal
probability over the total of the 20 numerical simulations
performed.

As for the evolution of a system with $g=\frac{0.14}{0.09}=1.555>1$
the system converges to the homogeneous solution $\tilde a=\tilde b=
1/(1+g)$ with small fluctuations about this value, i.e. the racemic state (III). All the results
shown appear for a wide range of fluctuation intensities
($\epsilon=10^{-2}-10^{-6}$).
%

\section{\label{sec:disc} Conclusions}

Internal particle density fluctuations automatically and inevitably
lead to chiral symmetry breaking in an extension \cite{Mason,GTV} of
the Frank model. The inclusion of this noise is carried out through
a first-principles technique which allows us to map the chemical
master equation to the corresponding coarse-grained Langevin
equations. This multiplicative noise is rigorously determined by
this procedure and is not an ad-hoc addition to the mean field
equations. A marked advantage of the fully stochastic model is that
no initial chiral seed nor external chiral field \cite{Avalos1} are
needed: full homochirality is achieved starting from a strictly
racemic initial condition. In the limit as $\epsilon \rightarrow 0$,
we would likely be sensitive to round-off errors, which also seem to
induce chiral symmetry breaking, as reported in \cite{Islas}. In two
dimensions, there is a second class of solutions (see Simulation 2
in Fig. 2) for which the final state consists of two enantiomeric
pure regions separated by a racemic interface or boundary. Regarding
the solutions displayed in Figs 1 and 2,  it is amusing to point out
that Frank, in conceiving of what may be expected to happen in
imperfectly mixed systems, predicted the formation of separate
``colonies" of the two kinds of enantiomers bounded by racemic
surfaces, with chiral homogeneity maintained within each such
colony. He also stated that any initial curvature in these
boundaries makes the boundary move towards the side from which it is
concave, a feature which we have also observed in our simulations.
The consequence of this is that any enclave of the one species will
eventually shrink, and the other survives; this is what we observe
in simulations of type 1 (see Fig 1). An exception to this rule
occurs ``when the two seas are connected by a strait. A different
species could survive in each sea, with the (racemic) boundary
running stably from cape to cape" \cite{Frank}. This is indeed what
happens in simulations of type 2 (see Fig.2)

An entirely different stochastic version of the Frank model was
presented in \cite{TGR}. Chiral amplification is also achieved
starting from a racemic initial state. There the noise is
incorporated in an ad-hoc fashion by assuming fluctuating kinetic
constants $k_i$.

We thank Dr. Vladik Avetisov for his interest in this work, as well
as Prof. Albert Moyano, and Prof. Josep M. Rib\'{o} for useful
discussions and comments. M.-P. Z. is supported by the INTA and the
research of D.H. is supported in part by funds from CSIC and INTA.

\end{document}